\documentclass[reprint,amsmath,amssymb,aps,showkeys]{revtex4-2}
\usepackage{graphicx}  % needed for figures
\usepackage{bm}        % for math
\usepackage{amssymb}   % for math
\usepackage{multirow}
\usepackage{indentfirst}
\usepackage{subfigure}
\usepackage{array}
\usepackage{xcolor}
\usepackage{mathtools}
\usepackage[none]{hyphenat}
\usepackage{hyperref}
\usepackage{bookmark}

\begin{document}
\title{Magnetic dipole-dipole transition for scintillation quenching}

\newcommand{\TsingHua}{\affiliation{Department~of~Engineering~Physics, Tsinghua~University, Beijing 100084, China}}
\newcommand{\HEPCenter}{\affiliation{Center for High Energy Physics, Tsinghua~University, Beijing 100084, China}}
\newcommand{\KeyLab}  {\affiliation{Key Laboratory of Particle \& Radiation Imaging (Tsinghua University), Ministry of Education, Beijing 100084, China}}

\author{Zhe~Wang}
\email[]{wangzhe-hep@mail.tsinghua.edu.cn}
\TsingHua\HEPCenter\KeyLab

\date{June 2, 2025}

\begin{abstract}
A magnetic dipole-dipole interaction is proposed as a mechanism for scintillation quenching.
The interaction rate follows $R^{-6}$ as the electric dipole-dipole interaction in F$\mathrm{\ddot{o}}$ster resonance energy transfer theory.
The proposed mechanism causes a long-range resonance energy transfer, and
the resonance condition is that the spins of donor and acceptor electrons both flip, and the energy level differences are the same.
The new mechanism is distinct to the known spin-orbit coupling induced intersystem crossing, and it can enhance the overall intersystem crossing rate.
When oxygen or organic molecules, including heavy elements, are dissolved in a liquid scintillator, these requirements are possible to be satisfied.
The proposal in the paper adds a new approach for scintillation quenching in liquid scintillators.
\end{abstract}

\maketitle
\section{Introduction}
Experimental research for neutrinoless double-beta decay is one of the frontiers of particle physics. The result will determine whether the neutrino is a Dirac particle or a Majorana particle~\cite{Majorana}.
One research approach is to dissolve tellurium in linear alkylbenzene (LAB) + 2,5-diphenyloxazole (PPO) based liquid scintillator
and to study the decay of tellurium-130~\cite{SNOPlusLS,SNOPlusTeBD}.
Besides loading tellurium in liquid scintillator, loading with other elements is also of great interest to physicists.
A comprehensive article about metal-loaded organic scintillators for neutrino physics can be found in Ref.~\cite{scintillator}.

These liquid scintillators usually include three components. The first one is the solvent, for instance, the LAB, which is dominant in mass and volume. The second component is the fluor, for instance, the PPO, which is at the level of only a few grams per liter. The last one is the chemicals for research, for instance, the organic tellurium compound for the neutrinoless double-beta decay study~\cite{SNOPlusTeBD}.
It is possible that other impurities are contained, such as oxygen.

The energy deposited by charged particles mainly causes the excitation and ionization of the solvent molecules.
The energy is transferred from the solvent molecules to the solute molecules primarily through non-radiative transfer processes, for instance, the electric dipole-dipole transitions, which is also known as F\"orster resonance energy transfer, FRET~\cite{Forster},
and the electric multipole-multipole resonance energy transfer suggested by Dexter~\cite{Dexter}.
The energy transfer may also occur through collision, for instance, the charge exchange effect~\cite{Dexter}.
The fluorescence of the solute molecules has a much longer attenuation length and detection efficiency for photomultiplier tubes.

However, when oxygen or molecules with heavy elements, for example, bromine, iodine, or lead, are dissolved in the solvent, a quenching effect always occurs~\cite{Birks}.
Depending on the structure and property of the involved molecules,
quenching can happen through the non-radiative energy transfer and collision~\cite{Forster, Dexter}, and the energy is dissipated to non-scintillating solutes.
Quenching can also happen through intersystem crossing induced by spin-orbit coupling~\cite{SpinOrbit}.
The intersystem crossing will cause the emission of phosphorescence instead of fluorescence.

In this work, a magnetic dipole-dipole transition mechanism is proposed as a scintillation quenching approach.
The mechanism can also introduce an intersystem crossing and cause the loss of fluorescence.
In Sec.~\ref{sec:Process}, the energy transfer process in normal scintillation and the proposed quenching processes are described.
Then the transition matrix is calculated in Sec.~\ref{sec:Matrix}.
The resonance condition and the influence of oxygen and heavy elements are presented in Sec.~\ref{sec:Resonance}.
The result is discussed in Sec.~\ref{sec:Discuss}, and a few experimental test proposals are raised in Sec.~\ref{sec:Exp}.
The paper is summarized in the last section.

\section{Scintillation and the proposed quenching effect}
\label{sec:Process}

\begin{figure}[h]
\centering
\includegraphics[width=8.5cm]{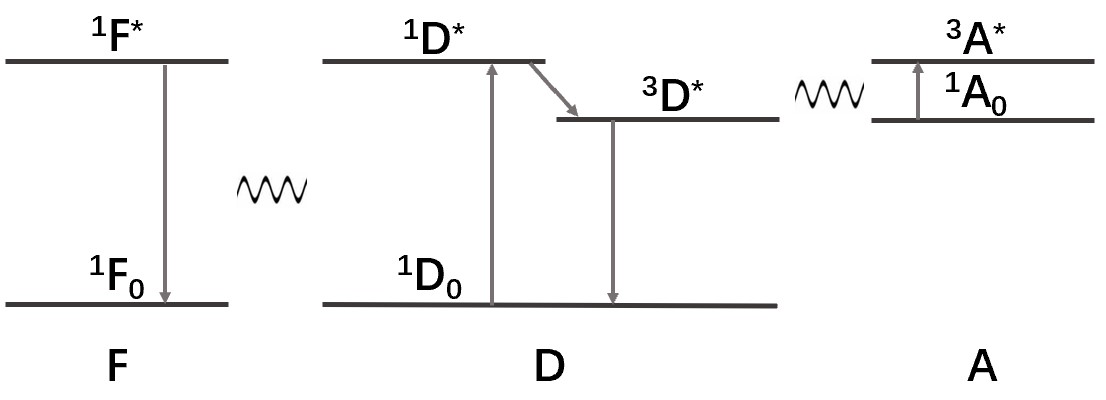}
\caption{Illustration of the electric dipole-dipole interaction from a donor to a fluor for fluorescence ($\mathrm{D} \to \mathrm{F}$) and
the proposed magnetic dipole-dipole interaction from a donor to an acceptor for quenching ($\mathrm{D} \to \mathrm{A}$).}
\label{fig:process}
\end{figure}

This paper is about the three-component liquid scintillators, such as LAB+PPO with other research chemicals.
The dominant and typical energy transfer process is plotted in Fig.~\ref{fig:process}, and
the proposed new quenching approach is also shown in the figure.
Other quenching mechanism is not the main topics of this paper.

In Fig.~\ref{fig:process} and the paper, the solvent molecule is denoted as energy donor, $\mathrm{D}$, and the fluor solute molecule is denoted as $\mathrm{F}$.
The donor's electron is first excited from its ground state ${^1\mathrm{D}_0}$ to its first singlet excited state ${^1\mathrm{D}^*}$.
The prescript 1 refers to the singlet state, in which the spins of the outmost electron pair of the donor molecule are anti-parallel.
The postscript 0 or * represents the ground or excited state, respectively.

The most abundant and dominant molecules in a liquid scintillator are solvent molecules, on which charged particles deposit most of their energy.
Donor's energy, when excited, is transferred to fluor solute molecules and, when the solute concentration is low, the energy transfer mechanism is
F$\mathrm{\ddot{o}}$ster resonance energy transfer (FRET), i.e.,~electric dipole-dipole interaction~\cite{Forster}.
This is the case, for example, in the liquid scintillator system of LAB+PPO.

The de-excitation of ${^1\mathrm{D}^*}$ to ${^1\mathrm{D}_0}$ will cause $\mathrm{F}$ to be excited from its ground state, ${^1\mathrm{F}_0}$, to the first excited singlet state, ${^1\mathrm{F}^*}$.
\begin{equation}
\label{eq:ElecDipoleTran}
{^1\mathrm{D}^*} +{^1\mathrm{F}_0} \to  {^1\mathrm{D}_0} +{^1\mathrm{F}^*}.
\end{equation}
The interaction is a pure electric dipole-dipole interaction, and the electron spin direction of D and F both remain unchanged.

${^1\mathrm{D}^*}$ can somehow decay to ${^3\mathrm{D}^*}$ spontaneously, for example, by intersystem crossing.
The prescript 3 refers to the triplet state, in which the spins of the outmost electron pair of the donor molecule are parallel.
The spin of the excited electron is flipped.
The previous $\mathrm{D} \to \mathrm{F}$ electric dipole-diploe interaction cannot happen because of spin forbidden,
and the fluorescence is quenched.
In turn, phosphorescence, ${^3\mathrm{D}^*} \to {^1\mathrm{D}_0}$, with a much longer emission time, will happen.

For the case involving heavy elements or oxygen in a liquid scintillator, a new quenching mechanism is proposed.
The quenching molecule is denoted as acceptor, $\mathrm{A}$.
A magnetic dipole-dipole interaction can happen and can stimulate or accelerate the quenching process, which is also shown in Fig.~\ref{fig:process}.
The flip of the electron spin from ${^1\mathrm{D}^*}$ to ${^3\mathrm{D}^*}$ causes a simultaneous flip of the spin of the acceptor's electron.
$\mathrm{A}$ is excited from its singlet state to a triplet state,
for example, from its ground state, ${^1\mathrm{A}_0}$, to the first excited triplet state, ${^3\mathrm{A}^*}$.
The quenching process is expressed as
\begin{equation}
\label{eq:MagDipoleTran}
{^1\mathrm{D}^*} +{^1\mathrm{A}_0} \to  {^3\mathrm{D}^*} +{^3\mathrm{A}^*}.
\end{equation}
The energy level difference between ${^1\mathrm{D}^*}$ and ${^3\mathrm{D}^*}$ must be equal to
the difference between ${^1\mathrm{A}_0}$ and ${^3\mathrm{A}^*}$, as required by the law of energy conservation.

\begin{figure}[h]
\centering
\includegraphics[width=5cm]{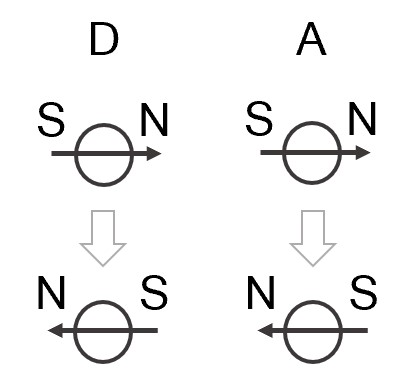}
\caption{Demonstration of magnetic dipole-dipole interaction with two magnets.
In the first row, the magnetic fields of the donor and acceptor electrons are aligned to have the lowest energy.
S and N are the magnetic south and north poles.
In the second row, when the spin of the donor electron flipped as well as its magnetic field, the acceptor electron's spin and magnetic field flipped too.}
\label{fig:MagneticDipole}
\end{figure}

The magnetic dipole-dipole interaction is like what happens to two adjacent magnets, and the flip of one magnet by 180 degrees will cause the neighbor one to flip if they are not quite far away from each other.
Figure~\ref{fig:MagneticDipole} demonstrates the interaction of the two magnets.

\section{Magnetic dipole-dipole interaction}
\label{sec:Matrix}
\begin{figure}[h]
\centering
\includegraphics[width=7cm]{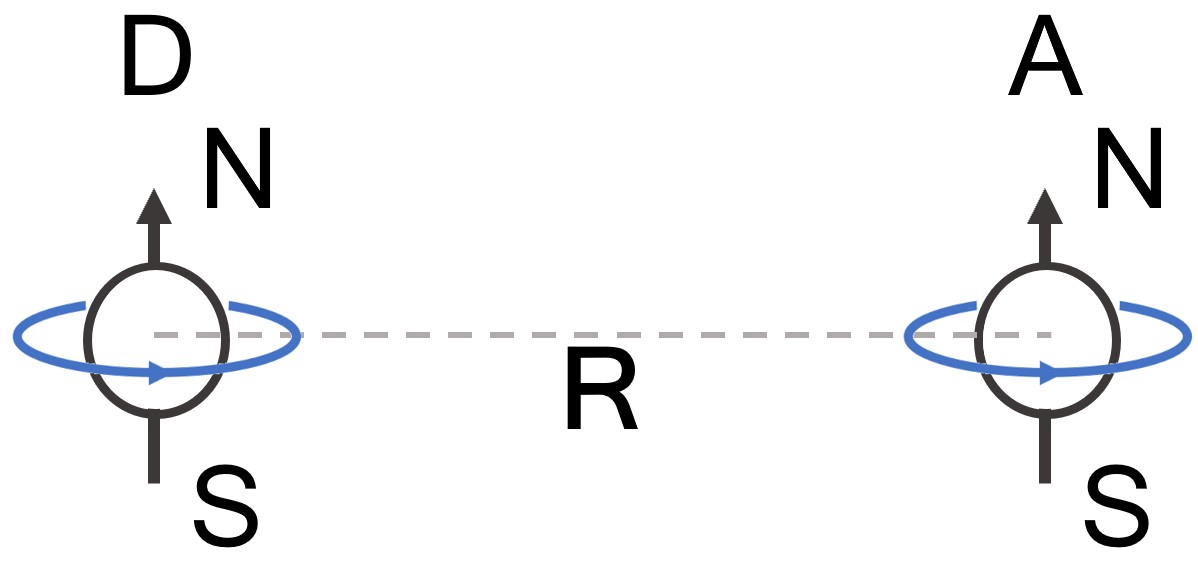}
\caption{Interaction of donor and acceptor. The spins of both donor and acceptor electrons are modeled by circular current.
The distance vector of the two electrons is $\mathbf{R}$.}
\label{fig:interaction}
\end{figure}

The magnetic dipole-dipole interaction is calculated.
The spin of an electron is interpreted as the result of a circular current, and
the magnetic interaction of two electrons is modeled as two circular currents, as shown in Fig.~\ref{fig:interaction}.

The magnetic field of the donor electron, $\mathbf{B}$, is
\begin{equation}
\label{eq:MagFld}
\begin{split}
\mathbf{B} &\approx -\frac{\mu_0 \mathbf{m}_{\mathrm{D}}}{4 \pi R^3} + \frac{3\mu_0 \mathbf{R} (\mathbf{m}_{\mathrm{D}} \cdot \mathbf{R})}{4 \pi R^5},
\end{split}
\end{equation}
where $\mu_0$ is the magnetic permeability, $\mathbf{m}_{\mathrm{D}}$ is the donor magnetic moment, and $\mathbf{R}$ is the distance  vector between the two electrons and $R=|\mathbf{R}|$. The result is obtained by assuming $R\gg r_0$, where $r_0$ is the geometrical radius of the circular current.
The result can be found in many popular college textbooks, for example, in Ref.~\cite{Electrodynamics,Electromagnetism}.

The magnetic energy of two electrons is modeled as the interaction of two circular currents.
The magnetic energy, $W$, is
\begin{equation}
\label{eq:EMag}
\begin{split}
W &= \mathbf{\mathbf{m}_{\mathrm{A}}} \cdot \mathbf{B},\\
  &= -\frac{\mathbf{\mu_0} \mathbf{m}_{\mathrm{A}} \cdot \mathbf{m}_{\mathrm{D}}}{4 \pi R^3}
     +\frac{3\mu_0 (\mathbf{m}_{\mathrm{A}} \cdot \mathbf{R}) (\mathbf{m}_{\mathrm{D}} \cdot \mathbf{R})}{4 \pi R^5},
\end{split}
\end{equation}
where $\mathbf{B}$ is the magnetic field caused by the donor electron, and $\mathbf{m}_{\mathrm{A}}$ is the magnetic moment of the acceptor electron.

Following the calculation method of FRET transition matrix element~\cite{Forster, PathtoFRET, QuantumChem},
the transition matrix element can be calculated as
\begin{equation}
\label{eq:MtxEle}
\begin{split}
H &= \langle \Psi_{{^3\mathrm{D}^*}} \Psi_{{^3\mathrm{A}^*}} | W | \Psi_{{^1\mathrm{D}^*}} \Psi_{{^1\mathrm{A}_0}} \rangle,
\end{split}
\end{equation}
where $\Psi_{{^3\mathrm{D}^*}}$, $\Psi_{{^3\mathrm{A}^*}}$, $\Psi_{{^1\mathrm{D}^*}}$, and $\Psi_{{^1\mathrm{A}_0}}$ are the wave functions for the four states in the magnetic dipole-dipole transition process in Eq.~\ref{eq:MagDipoleTran}.

According to Eq.~\ref{eq:EMag}, the total transition element, $H$, is calculated by including two terms, $H_1$ and $H_2$~\cite{QuantumChem}.
\begin{equation}
\label{eq:MtxElel}
\begin{split}
H = H_1 + H_2.
\end{split}
\end{equation}
The first term, $H_1$, is
\begin{equation}
\label{eq:MtxElel1}
\begin{split}
  H_1&= -\frac{\mathbf{\mu_0}}{4 \pi R^3}
  \langle \Psi_{{^3\mathrm{D}^*}} \Psi_{{^3\mathrm{A}^*}} | \mathbf{m}_\mathrm{A} \cdot \mathbf{m}_\mathrm{D} | \Psi_{{^1\mathrm{D}^*}} \Psi_{{^1\mathrm{A}_0}} \rangle, \\
     &= -\frac{\mathbf{\mu_0}}{4 \pi R^3}
      \langle \Psi_{{^3\mathrm{A}^*}} | \mathbf{m}_\mathrm{A} | \Psi_{{^1\mathrm{A}_0}} \rangle \cdot
      \langle \Psi_{{^3\mathrm{D}^*}} | \mathbf{m}_\mathrm{D} | \Psi_{{^1\mathrm{D}^*}} \rangle, \\
     &= -\frac{\mathbf{\mu_0}}{4 \pi R^3}  \mathbf{s}_\mathrm{A} \cdot \mathbf{s}_\mathrm{D}
\end{split}
\end{equation}
where $\mathbf{s}_\mathrm{A}$ and $\mathbf{s}_\mathrm{D}$ are their magnetic transition moments.
The second term of $H$ is
\begin{equation}
\label{eq:MtxElel2}
\begin{split}
H_2 = \frac{3\mu_0 (\mathbf{s}_\mathrm{A} \cdot \mathbf{R}) (\mathbf{s}_\mathrm{D} \cdot \mathbf{R})}{4 \pi R^5}.
\end{split}
\end{equation}

With Eq.~\ref{eq:MtxElel}, $H$ shows a dependency on $1/R^3$.
The transition rate, $k$, is
\begin{equation}
\label{eq:Rate}
k = \frac{2\pi}{\hbar}|H|^2 \rho \propto 1/R^{6},
\end{equation}
where $\rho$ is for the phase space and $k$ depends on $1/R^6$.

\section{Resonance condition}
\label{sec:Resonance}
The resonance condition for a magnetic resonance energy transfer (Eq.~\ref{eq:MagDipoleTran}) to occur is
spin matching and energy level gap matching between donor and acceptor in $\rho$.

Both the donor and acceptor electrons' spins flip simultaneously.
Paired electrons in the same state must have anti-parallel spin alignment, and the Pauli exclusion principle does not allow parallel arrangement and does not allow either electron to flip its spin.
If one electron is excited, then it is allowed to change its spin direction.
Another allowed situation is a lone electron. For example, the electron in a shell that is incompletely filled, or in a free radical.

The energy difference, $\Delta U$, of an electron before and after the spin flip in a magnetic field B is
\begin{equation}
\label{eq:UGap}
\begin{split}
\Delta U &= \Delta m_z B,\\
         &= \frac{q_e}{m_e} \hbar B,
\end{split}
\end{equation}
where $\Delta m_z$ is the change of magnetic moment in the direction of B, i.e.,~the $z$ direction,
$q_e$ and $m_e$ are the electron's charge and mass, respectively.
Electron spin projection is $\pm 1/2$ in any direction, and the change between them is always 1.
The result of $\Delta m_z$ is $({q_e}/{m_e})\hbar$~\cite{Feynman}.
The magnetic field $B$ can be understood as the electron's environmental magnetic field, contributed by other electrons.
The gap between ${^1\mathrm{D}^*}$ and ${^3\mathrm{D}^*}$ for benzene is $0.96~\rm{eV}$~\cite{benzene}.

\subsection{Oxygen}
The oxygen, $\rm{O_2}$, is paramagnetic, i.e.,~attracted to a magnet, because there are two unpaired electrons in its outermost orbital.
The ground state of $\rm{O_2}$ is a triplet~\cite{O2Level}. 
These electrons are ready to change their spin directions and stimulate the quenching.
The first excited state of $\rm{O_2}$ is a singlet, with the spin flipped for one electron.
The energy level of the first excited state is 0.977~eV~\cite{O2Level}.
The mole fraction solubility of oxygen in water is $2.5\times10^{-5}$~\cite{O2Solubility} for the room temperature and the standard atmosphere pressure,
and the mole fraction solubility of oxygen in benzene is $8.0\times10^{-4}$~\cite{O2inBenzene}.
It can be seen that the energy gap is consistent with the energy gap between the benzene's first excited singlet, ${^1\mathrm{D}^*}$, and the first excited triplet, ${^3\mathrm{D}^*}$, and, with the low solubility, a long-range interaction is expected.

\subsection{Heavy element molecules}
For organic molecules including heavy elements, two plausible guesses can be made.
For heavy elements, for example, chromium, iron, cobalt, and palladium, their inner electron shells are not completely filled.
Rare earth elements also have this property.
The property is inherited by the organic molecules, which include these elements.
These unpaired electrons are the target acceptor objects.
The second mechanism is slightly different.
The electron energy level difference between the ground state and the first excited state is very small when a heavy element is in the molecule.
The energy difference is close to or smaller than the energy difference between two vibration levels, so that
there are some fractions of electrons in their first excited states naturally, just due to thermal equilibrium.
These electrons are free to change their spin directions.
Certainly, the local magnetic field B is consistent with the donor's. 

\section{Discussion}
\label{sec:Discuss}
The electronic dipole-dipole interaction mechanism, FRET, is a long-range resonance energy transfer mechanism in
molecular systems and is the dominant mechanism in a liquid scintillator solution with dilute fluor concentration.
FRET is quite important in physics, chemistry, and biology~\cite{PathtoFRET}.
The energy of an electronic dipole-dipole system can be found in Ref.~\cite{PathtoFRET} too.
The transition rate of FRET is proportional to $1/R^6$.
The addition of PPO in LAB causes a much shorter lifetime of LAB due to FRET.
It decreases significantly from tens of ns to a few ns~\cite{LAB, LABPPO}.

Motivated by FRET, the magnetic dipole-dipole interaction mechanism is proposed in this work.
In a fermion-antifermion system, the magnetic dipole transition is quite feasible and known as M1-type decay~\cite{Electrodynamics, AtomicPhysics}.
For example, these can be found in the decay of charmonium systems~\cite{Charmonium}, mesons with a charm quark and an anti-charm quark,
$J/\psi \to \gamma + \eta_c$,
and $\psi(2S) \to \gamma + \eta_c'$.
The $\rm{J^{P}}$ (spin and parity) of $J/\psi$ and $\psi(2S)$ are both $1^-$, and that of $\eta_c$ and $\eta_c'$ are both $0^-$.
In view of particle physics, ${^1\mathrm{D}^*} \to {^3\mathrm{D}^*}$ processes occur naturally.

The magnetic dipole-dipole interaction is also a long-range resonance energy transfer mechanism.
The transition rate of magnetic dipole-dipole interaction is also proportional to $1/R^6$, which is the same as FRET.
The magnetic process will enhance the energy transfer in the same order of magnitude as FRET, causing a shorter lifetime and scintillation quenching for the donor, for example, LAB.

From the point of view of quantum electrodynamics, one virtual M1-type photon is emitted from the donor electron and absorbed by the acceptor electron~\cite{Electrodynamics, AtomicPhysics,QED}.
The parity of the M1-type photon is +1.
The electric dipole radiation is mediated by a virtual E1-type photon, which has a parity of -1.
Parity conservation requires that there is no mixing between these two processes.
The proposed magnetic dipole-dipole interaction is different from FRET.

\section{Possible experimental tests}
\label{sec:Exp}
To avoid the quenching induced by the magnetic dipole-dipole transition, some study methods for oxygen gas~\cite{oxygen} can be applied. 
It is important to first identify whether there is any absorption band around 1240 nm, i.e. 1 eV.
Then one needs to identify whether the absorption band is from a spin-flip transition or a spin-conserving transition. 
If it is a spin-flip transition, it will be a target acceptor for the magnetic dipole-dipole transition.

Xenon is a good example, which has a high proton number, Z, and a moderate solubility in organic liquid scintillator, but a minor quenching effect. 
According to the reported information given by the KamLAND-zen experiment~\cite{KamLAND-Zen}, 754 kg of xenon has been dissolved in liquid scintillator, and the mass fraction is 3.13\%; no serious scintillation quenching has been reported.
Checking the first few excited energy levels, one can find that the energy gap between the ground state and the first excited state has reached 8.3 eV~\cite{nist}. 
It does not satisfy the resonance condition of electric or magnetic dipole-dipole transitions, so 
there is no serious quenching induced by electric or magnetic dipole-dipole transitions.

One can also think of using the known instrument, electron-paramagnetic resonance, EPR, to examine each interested liquid scintillator component and to search for free radicals, i.e.,~lone electrons.
To see the free radical signals of LAB, the sample needs to be illuminated by ultraviolet light to excited states.
One can also use the known organic with free radicals to check whether there is any enhanced quenching.
But there are two difficulties in these tests.
The first difficulty is to eliminate the interference of other quenching factors.
The second difficulty is that the free radicals in the donor and acceptor may not satisfy the resonance condition,
i.e.,~the magnetic fields of Eq.~\ref{eq:UGap} for the lone electrons of the donor and acceptor do not match.

The known EPR can be extended to study the resonance condition.
The current EPR devices apply two magnetic fields to the testing sample.
One static magnetic field, $B_0$, is around 1 Tesla. The other one is an alternating magnetic field around tens giga Hz orthogonal to $B_0$. The current EPR instrument resonance signals occur at $B_0$, which completely covers the atom's own magnetic field.
The extension is to turn off the static magnetic field, i.e.,~with $B_0=0$, and to search for EPR resonance signals.
The new measurement will tell us what the atom's magnetic field in Eq.~\ref{eq:UGap} is or what the range is.

\section{Summary}
\label{sec:Summary}
In this paper, a magnetic dipole-dipole interaction is proposed as a scintillation quenching mechanism.
The interaction rate also follows $R^{-6}$ as FRET and can cause a long-range resonance energy transfer.
The resonance condition is that the spins of donor and acceptor electrons both flip, and the energy level differences are the same.
The new mechanism is distinct from the known spin-orbit coupling induced intersystem crossing, and it can enhance the overall intersystem crossing rate.
When oxygen or heavy elements are dissolved in a liquid scintillator, these requirements are easier to satisfy.
The proposal in the paper adds a new theoretical approach of quenching in a liquid scintillator solution.

\section{Acknowledgement}
This work is supported in part by
the National Natural Science Foundation of China (No.~12141503),
the Ministry of Science and Technology of China (No.~2022YFA1604704),
the Key Laboratory of Particle \& Radiation Imaging (Tsinghua University),
and the CAS Center for Excellence in Particle Physics (CCEPP).
I also like to thank Minfang Yeh and Ye Liang for helping me to digest many chemical concepts, and
to thank Qing Wang and Zhicai Zhang for helping me to pick up a few electromagnetism conclusions.

% BibTeX users please use
%\bibliographystyle{unsrt}
\bibliography{MagneticDipole}

\end{document}